\documentclass[conference]{IEEEtran}
\IEEEoverridecommandlockouts
\usepackage{algorithm}
\usepackage{subfigure}
\usepackage{hyperref}
\usepackage{multirow}
\usepackage{balance}

\usepackage{cite}
\usepackage{amsmath,amssymb,amsfonts}
\usepackage{algorithmic}
\usepackage{graphicx}
\usepackage{textcomp}
\usepackage{xcolor}
\def\BibTeX{{\rm B\kern-.05em{\sc i\kern-.025em b}\kern-.08em
    T\kern-.1667em\lower.7ex\hbox{E}\kern-.125emX}}
\begin{document}

\title{
%DPI-TTS: Directional patch interaction based on temporal and spectral sequences for fast-converging and fine-grained style control in Text-to-Speech
DPI-TTS: Directional Patch Interaction for Fast-Converging and Style Temporal Modeling in Text-to-Speech

}

\author{
    Xin Qi\textsuperscript{1}, 
    Ruibo Fu\textsuperscript{1,\textsuperscript{\dag}},
    Zhengqi Wen\textsuperscript{2},
    Tao Wang\textsuperscript{1}, 
    Chunyu Qiang\textsuperscript{1}, 
    Jianhua Tao\textsuperscript{2},
    Chenxing Li\textsuperscript{3},
    Yi Lu\textsuperscript{1},\\
    Shuchen Shi\textsuperscript{1},
    Zhiyong Wang\textsuperscript{1},
    Xiaopeng Wang\textsuperscript{1},
    Yuankun Xie\textsuperscript{1},
    Yukun Liu\textsuperscript{1},
    Xuefei Liu\textsuperscript{1},
    Guanjun Li\textsuperscript{1}
    \vspace{1.5mm}
    \\
    \IEEEauthorblockA{
    \textsuperscript{1}Institute of Automation, Chinese Academy of Science, Beijing, China
    }
    \IEEEauthorblockA{
    \textsuperscript{2}Beijing National Research Center for Information Science and Technology, Tsinghua University, Beijing, China
    }
    \IEEEauthorblockA{
    \textsuperscript{3}AI Lab, Tencent, Beijing, China
    }
    \IEEEauthorblockA{qixin221@mails.ucas.ac.cn, ruibo.fu@nlpr.ia.ac.cn}
    \thanks{\textsuperscript{\dag} Corresponding author.}
    \thanks{Our demo and code are available at \href{https://7xin.github.io/DPI-TTS/}{https://7xin.github.io/DPI-TTS/}.}
}

\maketitle

\begin{abstract}
%近年来，语音扩散模型发展迅速。除了广泛使用的U-net架构外，基于变压器的模型，如DiT（扩散变压器）也受到了关注。
%但是，目前的DiT语音模型是将梅尔频谱图作为一般图像处理，忽略了语音的声学属性。
%To address these limitations, 我们提出了基于DiT的Directional Patch Interaction方法被称作DPI-TTS,在快速训练的同时还不损失精度。
%令人惊奇的是，DPI-TTS的从低频到高频、逐帧步进式推理的方式，更符合声学属性，能提升自然度。
%除此之外，细粒度的风格时序建模方法被提出，进一步提升DPI-TTS的说话人风格相似度。
%实验结果表明，我们方法训练速度快了近一倍，表现好于基线。
In recent years, speech diffusion models have advanced rapidly. Alongside the widely used U-Net architecture, transformer-based models such as the Diffusion Transformer (DiT) have also gained attention. However, current DiT speech models treat Mel spectrograms as general images, which overlooks the specific acoustic properties of speech. 
To address these limitations, we propose a method called Directional Patch Interaction for Text-to-Speech (DPI-TTS), which builds on DiT and achieves fast training without compromising accuracy. Notably, DPI-TTS employs a low-to-high frequency, frame-by-frame progressive inference approach that aligns more closely with acoustic properties, enhancing the naturalness of the generated speech. Additionally, we introduce a fine-grained style temporal modeling method that further improves speaker style similarity. Experimental results demonstrate that our method increases the training speed by nearly 2 times and significantly outperforms the baseline models.
\end{abstract}

\begin{IEEEkeywords}
Acoustic Properties, fast-converging, Directional Interaction, Text-to-Speech
\end{IEEEkeywords}

\section{Introduction}

%1.主流TTS发展，现在的有的DiT改进，相对传统diff的好处，给语音带来了那些发展。（2~3句）。
%2.目前diff的主要TTS工作 -> DiT TTS -->主要的不足1）把频谱当作为特殊图像建模
%3.从音频而言，频谱和图像还是有差别 -> (对于语音频谱建模而言，每一帧只跟相邻的信息是相关的，从物理特性上，后一帧只和前一帧相关)
%4.我们想 针对语音的特性，构建频谱，
    %1）更快的推理速度。
    %2）基于声学原理的有向推理
    %3）细粒度的风格建模。

%\begin{figure}[h]
%	\centering
%	\subfloat[]{\includegraphics[width=0.5\linewidth]{fig/Loss_comparison.png}%
%		\label{fig_first_case}}
%	\subfloat[]{\includegraphics[width=0.5\linewidth]{fig/Loss-Epoch.png}
%		\label{fig_first_case}}
%	\caption{The loss curves of our method and the baseline on the LJSpeech dataset. To reach the same level, our method requires less time. At the same time points, our method demonstrates better performance.}
%	\label{Lc}
%\end{figure}

%Text-to-Speech(TTS)\cite{shen2018natural,ren2019fastspeech,ren2020fastspeech}是将给定文本建模为自然语音的任务，具备很高的应用价值。
%在开发的众多方法中，语音扩散模型因其生成高表达性和自然发音语音的能力而受到广泛关注。
%时至今日，语音扩散模型经过各种各样的改进，也渐渐发展出了不同的派别。
Text-to-Speech (TTS) \cite{shen2018natural,ren2019fastspeech,ren2020fastspeech} is the task of generating natural speech from a given text, offering significant application value across various domains. 
Among the many approaches developed for TTS, speech diffusion models have garnered considerable attention due to their impressive capabilities in producing highly expressive and natural speech.
These models have undergone numerous improvements in recent years and evolved into distinct variants, each focusing on different aspects of TTS.

%在Diffusion Decoder中使用k个全局DiT块来补充全局细节，比如基频等，再用（N-k）个有向DiT块进行精确建模。在有向DiT块中，第一步添加时域位置信息，从语音帧的角度出发与说话人风格信息融合；第二步添加时域和频域的位置信息，每一个patch仅与周围三个块，即前一帧和更低频的部分进行注意力计算。
\begin{figure*}[ht]
  \centering
  \includegraphics[width=\linewidth]{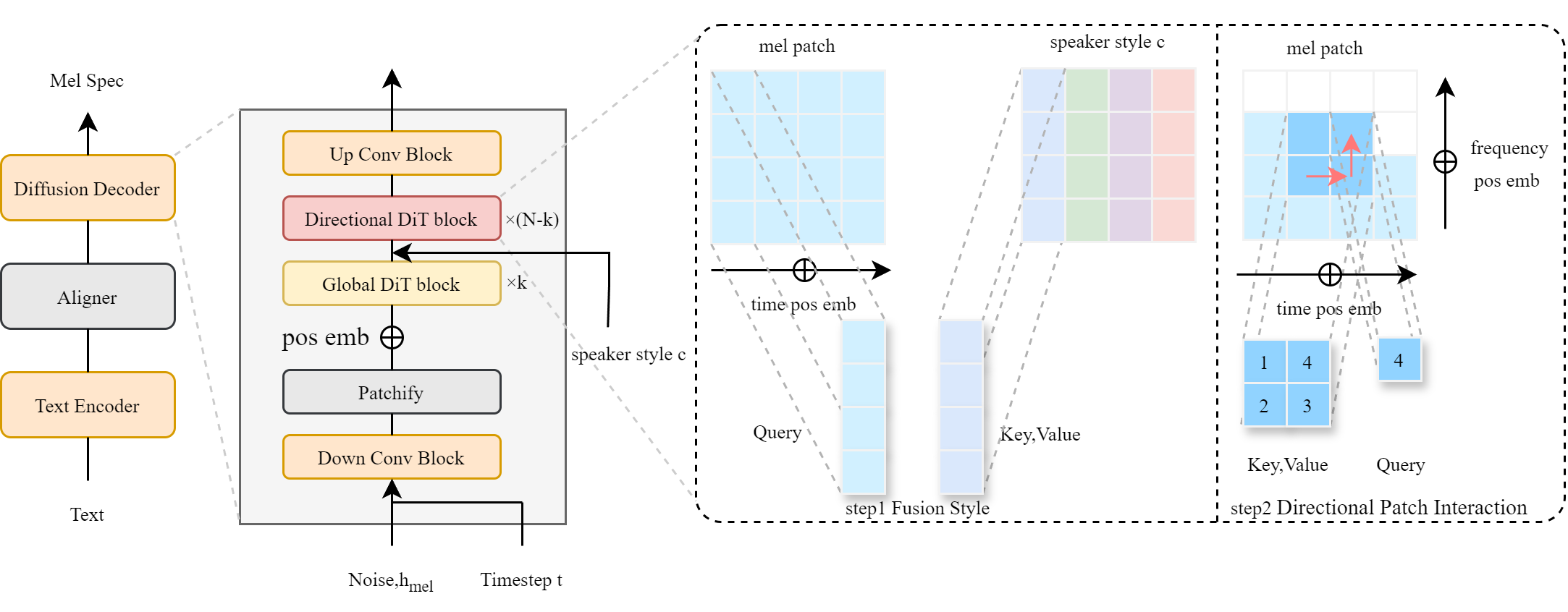}
  \caption{An overview of our method. In the Diffusion Decoder, \( k \) global DiT blocks are used to supplement global details such as fundamental frequency, followed by \((N-k)\) directional DiT blocks for precise modeling. In the directional DiT blocks, the first step involves adding time positional embeddings and integrating them with speaker-style information from a temporal perspective. The second step incorporates both time and frequency domain positional embeddings, where each patch computes attention only with three surrounding patches: the previous frame and lower-frequency components.}
  \label{overview}
\end{figure*}

%U-Net\cite{ronneberger2015u}作为语音扩散模型最主流的框架之一，在声学模型和声码器领域都产生了很多优秀的工作。
%Diff-TTS \cite{jeong2021diff} 和 Grad-TTS \cite{popov2021grad} 通过改进TTS扩散过程提升效果。
%Prodiff \cite{huang2022prodiff} 和 Comospeech \cite{ye2023comospeech}实现了加速推理。
%NaturalSpeech \cite{tan2024naturalspeech} 和NaturalSpeech2 \cite{shen2023naturalspeech}实现了接近人类的语音质量。
%DiffWave \cite{kong2020diffwave} 引入了一种通用的扩散声码器。
%FastDiff \cite{huang2022fastdiff} 和 InferGrad \cite{chen2022infergrad} 减少了声码器推理的迭代次数。
%SpecGrad \cite{koizumi2022specgrad} 和 PriorGrad \cite{lee2021priorgrad} 则提升了声码器性能
%随着研究不断深入，有研究者发现U-Net的归纳偏差对Diffusion模型的性能并不是至关重要的\cite{peebles2023scalable}。
%如果能用Transformer\cite{vaswani2017attention}结构取代U-Net，那么现有的transformer提技巧都可以迁移过来让扩散模型受益。
U-Net \cite{ronneberger2015u} is one of the most prominent frameworks in speech diffusion models and has contributed to significant advancements in acoustic modeling and vocoder development. 
Notable models such as Diff-TTS \cite{jeong2021diff} and Grad-TTS \cite{popov2021grad} have optimized the TTS diffusion processes, leading to enhanced performance. 
Models like Prodiff \cite{huang2022prodiff} and Comospeech \cite{ye2023comospeech} have achieved faster inference times, while NaturalSpeech \cite{tan2024naturalspeech} and NaturalSpeech2 \cite{shen2023naturalspeech} have reached near-human speech quality. 
DiffWave \cite{kong2020diffwave} introduced a flexible diffusion vocoder, and models such as FastDiff \cite{huang2022fastdiff} and InferGrad \cite{chen2022infergrad} have reduced the number of iterations required for vocoder inference. 
Additional improvements in vocoder performance have been made by SpecGrad \cite{koizumi2022specgrad} and PriorGrad \cite{lee2021priorgrad}.
As research has advanced, it has been observed that the inductive bias of U-Net is not a critical factor in the performance of diffusion models \cite{peebles2023scalable}. 
This finding opens the possibility of replacing U-Net with Transformer \cite{vaswani2017attention} architectures, enabling the transfer of existing Transformer enhancement techniques to diffusion models, potentially leading to significant performance gains.

%DiT\cite{peebles2023scalable}作为以transformer为框架的的新型扩散模型吸引了诸多关注，最近的DiT-TTS工作都展现出了更好的性能和更精细的建模能力。
%U-DiT\cite{jing2023u}在单说话人数据集LJSpeech\cite{ljspeech17}上达到了最先进的性能。
%DiTTo-TTS\cite{lee2024ditto}通过跨注意力机制和预测语音表示的总长度来改善文本与语音对齐。
%Dex-TTS\cite{park2024dex}通过将风格信息分为time-invariant and time-variant categories融合提升语音风格表现力。
%尽管目前的DiT的TTS工作取得了不错的表现，但他们都有一个局限性就是忽略了语音的声学属性，将梅尔频谱的图像作为一般图像来处理。
DiT \cite{peebles2023scalable}, a novel diffusion model based on Transformer architectures, has garnered significant attention due to its enhanced modeling capabilities.
Recent work on DiT-TTS has demonstrated improved performance and more precise modeling of speech characteristics. 
Notably, U-DiT \cite{jing2023u} achieved state-of-the-art (SOTA) performance on the single-speaker dataset LJSpeech \cite{ljspeech17}. 
DiTTo-TTS \cite{lee2024ditto} further improves TTS alignment through cross-attention mechanisms and by predicting the total length of speech representations. 
Dex-TTS \cite{park2024dex} advances speech style expressiveness by incorporating style information, categorized into time-invariant and time-variant components. 
Despite these advancements, current DiT-based TTS models share a common limitation: they treat Mel spectrograms as general images, neglecting the intrinsic acoustic properties of speech.

%一般图像则更注重整个图片每一部分的统一性和关联性，需要全局计算或全局信息作为引导。
%但对语音来说，这样做会导致结果过于平滑。
%从声学讲，梅尔频谱\cite{hu2024mel,zhou2024mel,xu2024convconcatnet,tian2023diffusion,qian2022cmelgan}有着更强的时序相关性，每一帧与前一帧的联系最为密切。不同的频率包含的能量也不同，人耳更容易感受低频。
%因此，我们基于DiT提出了一种patch定向交互的方法，被称作DPI-TTS:Directional Patch Interaction for Fast-Converging and Style Temporal Modeling in Text-to-Speech。
%因此，我们提出了 DPI-TTS:Directional Patch Interaction for Fast-Converging and Style Temporal Modeling in Text-to-Speech。
%根据语音声学属性，DPI-TTS将mel patch通过切分和拼接，找到每个patch对应交互的前一帧和低频部分计算注意力实现Directional Patch Interaction。这种方法在不损失精度下提高了训练速度，同时提高自然度和时序一致性。
%同时DPI-TTS通过交叉注意力将风格信息按时间顺序逐步融入mel patch中，提升Directional Patch Interaction下的说话人相似度。
%实验结果表明，DPI-TTS在提升近一倍的训练速度情况下，表现优于基线。

General image processing often emphasizes the uniformity and interconnectedness of all parts, typically requiring global computation or information as guidance. 
However, applying this approach to speech can result in overly smooth outputs that fail to capture the nuanced acoustic properties of speech signals. 
From an acoustic perspective, Mel spectrograms \cite{hu2024mel,zhou2024mel,xu2024convconcatnet,tian2023diffusion,qian2022cmelgan} exhibit strong temporal correlations, where each frame is most closely related to its preceding frame. 
Additionally, different frequencies contain varying energy levels, with the human ear being more sensitive to lower frequencies.
Therefore, we propose a novel method called DPI-TTS (Directional Patch Interaction for Fast-Converging and Style Temporal Modeling in Text-to-Speech), which leverages patch directional interaction based on DiT. 
DPI-TTS segments Mel spectrograms into patches and computes attention by focusing on the interaction between each patch, its preceding frame, and low-frequency components, realizing directional patch interaction. 
This approach enhances training speed without compromising accuracy and improves synthesized speech's naturalness and temporal consistency. 
Furthermore, DPI-TTS incrementally incorporates style information into Mel patches chronologically through cross-attention, enhancing speaker similarity under directional patch interaction. 
Experimental results demonstrate that DPI-TTS outperforms baseline models and increases the training speed by 2 times.

%我们的贡献如下：
Our contributions are as follows:
\begin{itemize}
%首先，DPI-TTS在不损失精度下提升了训练速度。Directional Patch Interaction将每个patch的注意力从整个梅尔频谱转变为局部分量。
%其次，DPI-TTS提升了提升语音时间一致性和自然度。DPI-TTS根据声学属性从低频到高频，逐帧步进式的推出。
%最后，DPI-TTS提供了一种细粒度的说话人风格时序建模方法，提升说话人相似度。

\item Firstly, DPI-TTS significantly improves training speed without compromising accuracy. By utilizing Directional Patch Interaction, the model shifts attention from the entire Mel spectrogram to localized components.

\item Secondly, DPI-TTS enhances temporal consistency and naturalness in speech by performing progressive inference from low to high frequencies, frame by frame, in alignment with acoustic properties.

\item Finally, DPI-TTS provides a fine-grained speaker-style temporal modeling method, enhancing speaker similarity.

\end{itemize}

\section{Method}
%DPI-TTS包含了一个由8层具有多头注意力机制(MHSA)的Transformer\cite{vaswani2017attention}块的文本编码器，编码器中加入相对位置信息relative position embedding(RoPE)\cite{su2024roformer}.同时还有一个基于卷积的时长预测器 (Duration Predictor, DP)\cite{kim2021conditional}aligner，将文本信息映射到用于扩散解码器中的初始mel频谱图表示h_mel的帧上，通过单调对齐搜索（Monotonic Alignment Search, MAS）算法进行训练。最后是一个Diffusion Decode，r主要包括一个Down Conv Block来减少计算量，一个Patchify将梅尔频谱图像分成patch。k个Global DiT block，(N-k）个Directional DiT block，最后接一个up Conv Block恢复特征。Global DiT block负责建模音频的全局信息，例如基频等。Directional DiT block负责风格时序建模，以及从声学属性角度上对语音的patch进行directional Interaction。
DPI-TTS features a text encoder with 8 Transformer layers using multi-head self-attention (MHSA) \cite{vaswani2017attention} and relative position embeddings (RoPE) \cite{su2024roformer}. It includes a convolution-based Duration Predictor (DP) \cite{kim2021conditional} aligner that maps text to the initial Mel spectrogram frames \( h_{mel} \). 
The Diffusion Decoder comprises a Down Conv Block, a Patchify module for segmenting the Mel spectrogram into patches, \( k \) Global DiT blocks, \( N-k \) Directional DiT blocks, and an Up Conv Block for feature restoration. Global DiT blocks capture global speech information like pitch, while Directional DiT blocks handle style temporal modeling and directional interactions on mel patches.

\subsection{Directional Patch Interaction}

%语音信号随时间动态变化，不同时间点的语音信号携带的信息是不同的，例如语音中的停顿、重音、节奏和韵律等因素都具备时序性。
%同时，语音低频和高频部分包含的能量也不同，人耳对低频部分感知更加敏感。
%让每个mel patch与前一帧和低频部分关联而非与整个频谱关联，有利于保持这些时序相关的动态变化，能增强低频信息的表现力，加强局部细节建模能力。

%具体来说，我们首先计算每个梅尔频谱块的query，key和value值。此时query，key和value形状为(b,h,w,d)。
%然后将query，key和value都预先插入一个维度变为(b,h,w,1,d)，其中b是batch大小，h是patch块在频域轴上的patch个数，w是patch块在时域轴上的patch个数，d是特征维度。
%再将key和value的最后一行和第一列与key和value拼接在一起，形状变为(b,h+1,w+1,1,d).
%紧接着用大小为(h,w)的窗口切分并将结果在第3个维度拼接,此时形状变为(b,h,w,4,d)。
%实现细节详见算法\ref{catlayer}.
%最后将所有patch展平，进行注意力计算，返回形状为(b,hw,1,d)的结果。
Speech signals change dynamically over time, with the information they convey varying at different moments.
Factors such as pauses, emphasis, rhythm, and prosody all possess distinct temporal properties in speech.
Furthermore, the energy distribution between low and high-frequency components of speech varies, with the human ear being more sensitive to lower frequencies. 
By associating each Mel patch with its preceding frame and low-frequency components rather than the entire spectrum, this approach preserves dynamic temporal changes, improves low-frequency information representation, and enhances local details' modeling.

Specifically, we first compute the query, key, and value for each Mel spectrogram image patch, with initial shapes of \((b, h, w, d)\), where \(b\) denotes the batch size, \(h\) is the number of patches along the frequency axis, \(w\) is the number of patches along the time axis, and \(d\) represents the feature dimension.
Next, an additional dimension is inserted into the query, key, and value, altering their shapes to \((b, h, w, 1, d)\). The last row and the first column of the key and value are then concatenated with the key and value, resulting in shapes of \((b, h+1, w+1, 1, d)\).
Subsequently, the concatenated key and value are split using a window of size \((h, w)\), and the results are concatenated along the third dimension, changing the shape to \((b, h, w, 4, d)\). For implementation details, please refer to Algorithm \ref{catlayer}.
Finally, all patches are flattened, and attention computation is performed, yielding a result with the shape \((b, hw, 1, d)\).

% \begin{algorithm}[tb]
%     \caption{Directional DiT block}
%     \label{ddb}
%     \textbf{Input}:mel patches: $x$, speaker emb: $c$\textbackslash\textbackslash$x_{shape}=(b,hw,d)$, $c_{shape}=(b,d)$\\
%     \textbf{Parameter}: $b$: batch size, $hw$: patch size, $d$: hidden dim\\
%     \textbf{Output}:$x$\textbackslash\textbackslash$x_{shape}=(b,hw,d)$
%     \begin{algorithmic}[1] %[1] enables line numbers      
%         \STATE $c.unsqueeze(1).repeat(1,x.shape[1],1)$
%         \STATE $x.reshape(bw,h,d),c.reshape(bw,h,d)$
%         \STATE $q_1=W^1_q(x),k_1=W^1_k(c),v_1=W^1_v(c)$
%         \STATE $x=attention(q_1,k_1,v_1)$
%         \STATE $q_2,k_2,v_2=W_{qkv}(x).split(3)$\textbackslash\textbackslash$shape=(b,hw,d)$
%         \STATE $together(q_2,k_2,v_2).reshape(b,h,w,1,d)$
%         \STATE $k_2=DirectPrepLay(k_2),v_2=DirectPrepLay(v_2)$
%         \STATE $flatten(q_2,k_2,v_2)$
%         \STATE $x=attention(q_2,k_2,v_2)$
%         \STATE \textbf{return} $x$
%     \end{algorithmic}
% \end{algorithm}

\subsection{Fine-Grained Speaker Style Temporal Modeling}
%得益于DiT语音合成模型将梅尔频谱分为多个patch的做法，DPI-TTS可以对风格进行细粒度的时序控制。
%常规方法通常是直接给出风格id或者从参考语音中提取的全局风格信息来融入整段语音中，容易导致风格表现过于平滑，难以实现这种显示的精确操作。
%并且常规方法下DPI-TTS的每个patch只能看到一部分风格信息，影响风格表达。
%因此我们针对DPI-TTS的特点，按照时间顺序，依次与风格信息融合。以每个时刻的所有patch为整体，保证高低频之间风格表现的一致性。

%如图\ref{overview}所示。
%DPI-TTS将mel patches作为查询，说话人风格作为key和value。
%为了提升模型对风格的时序感知能力，再向mel patches中加入时域位置信息。
%最后按照时间维度，对每一组patch做交叉注意力计算，将风格特征融合到语音当中。
Thanks to the DiT-based TTS model’s approach of dividing the Mel spectrogram into multiple patches, DPI-TTS enables fine-grained temporal control of style.
Conventional methods typically rely on using a style ID or incorporating global style information extracted from reference speech across the entire speech. 
This often leads to overly smooth style expression and makes precise control challenging.
In these conventional approaches, each patch in DPI-TTS only accesses partial style information, which can negatively impact style expression. 
To address this, DPI-TTS integrates style information sequentially over time, treating all patches at each time point as a cohesive unit. 
This ensures consistent style representation across high and low frequencies, aligning with the model’s design characteristics.

As shown in Figure \ref{overview},
DPI-TTS uses mel patches as queries, with speaker style as keys and values.
To enhance the model’s temporal awareness of style, time positional information is added to the mel patches. 
Finally, cross-attention is computed along the time dimension for each group of patches, integrating the style features into the speech.

\section{Experiment}

\begin{algorithm}[tb]
    \caption{DirectionalPatchInteraction}
    \label{catlayer}
    \textbf{Input}:$x$\textbackslash\textbackslash$x_{shape}=(b,h,w,1,d)$\\
    \textbf{Parameter}: $b$: batch size, $hw$: patch size, $d$: hidden dim\\
    \textbf{Output}:$x$\textbackslash\textbackslash$x_{shape}=(b,h,w,4,d)$\\
    \vspace{-3ex}
    \begin{algorithmic}[1] %[1] enables line numbers      
        \STATE $cat_x=concat([x, x[:, -1:]], dim=1)$
        \STATE $cat_x = concat([cat_x[:, :, 0:1], cat_x], dim=2)$
        \STATE $x = concat([x, $\\
                            $cat_x[:, :x.shape[1], :x.shape[2]],$\\
                            $cat_x[:, 1:, :x.shape[2]], $\\
                            $cat_x[:, :x.shape[1], 1:]], dim=3)$
        \STATE \textbf{return} $x$
    \end{algorithmic}
\end{algorithm}

\subsection{Experiment Setup}
\subsubsection{Dataset}
%为了评估我们的方法，我们使用了英文单说话人数据集LJSpeech\cite{ljspeech17}和英文多说话人VCTK\cite{Veaux2017CSTRVC}数据集。
%其中VCTK数据集包含了approximately 400 utterances per 109 speakers.
%我们将每个数据集按照70%,15%,和15%的比例划分训练，验证和测试集。
To evaluate our method, we used the English single-speaker dataset LJSpeech \cite{ljspeech17} and the English multi-speaker VCTK dataset \cite{Veaux2017CSTRVC}. The VCTK dataset contains approximately 400 utterances per 109 speakers. Each dataset was split into training, validation, and test sets in a ratio of 70\%, 15\%, and 15\%, respectively.
\subsubsection{Baselines}
%为了对比，我们设置了以下几个基线：
%1)参考语音
%2)VITS,一个基于VAE\cite{kingma2013auto}和flow\cite{rezende2015variational}的多说话人TTS模型
%3)Grad-TTS，基于U-Net框架的语音合成扩散模型，用HiFi-GAN作为声码器。
%4)Dex-TTS，基于DiT框架的语音合成扩散模型，用HiFi-GAN作为声码器。
For comparison, we set the following systems as baselines: 
\textbf{1) GT}, the ground-truth.
\textbf{2)VITS}, a multi-speaker TTS model based on VAE \cite{kingma2013auto} and flow \cite{rezende2015variational}.
\textbf{3)Grad-TTS}, a TTS diffusion model based on the U-Net framework, using HiFi-GAN as the vocoder.
\textbf{4)Dex-TTS}, a TTS diffusion model based on the DiT framework, using HiFi-GAN as the vocoder.

\subsubsection{Implementation Details}
Training uses 1000 epochs for VCTK and 2000 epochs for LJSpeech. 
Parameters include a batch size of 32, patch size of 7, NN of 4, hidden dimension of 64, diffusion steps of 50, and kk of 2. 
Experiments were run on an NVIDIA 3090 GPU.
\subsubsection{Evaluation Metrics}
%对于客观评估，我们考虑了WER和COS。WER是单词错误率，由预训练的WAV2VEC模型识别语音，与文本计算单词错误率得到。
%COS由预训练的说话人识别模型识别合成语音和参考语音的说话人特征计算余弦相似度得到,并放大100倍。
%对于主观评估，我们采用自然度平均意见得分MOS-N和说话人相似度平均意见得分MOS-S，评分范围为1~5分。
%我们聘请了至少20位预先经过培训的评审人员，评估每个说话人至少50条语音。
% For objective evaluation, we used Word Error Rate (WER\%) and Cosine Similarity (COS). WER is derived from a pre-trained Wav2Vec model for speech recognition, while COS is calculated from the cosine similarity between synthesized and reference speaker features, scaled by 100.
% For subjective evaluation, we assessed naturalness (MOS-N) and speaker similarity (MOS-S) on a 1 to 5 scale. At least 20 evaluators rated a minimum of 50 samples per speaker.
For objective evaluation, we utilized Word Error Rate (WER\%) and Cosine Similarity (COS). WER was obtained using a pre-trained Wav2Vec speech recognition model. At the same time, COS was calculated as the cosine similarity between synthesized and reference speaker features extracted from a pre-trained speaker recognition model, scaled by 100. For subjective evaluation, we assessed naturalness (MOS-N) and speaker similarity (MOS-S) on a scale from 1 to 5. At least 20 evaluators rated a minimum of 50 samples per speaker.
\begin{figure*}[ht]
      \centering
      \subfigure[Loss curve of baseline model]{
           \includegraphics[scale=0.38]{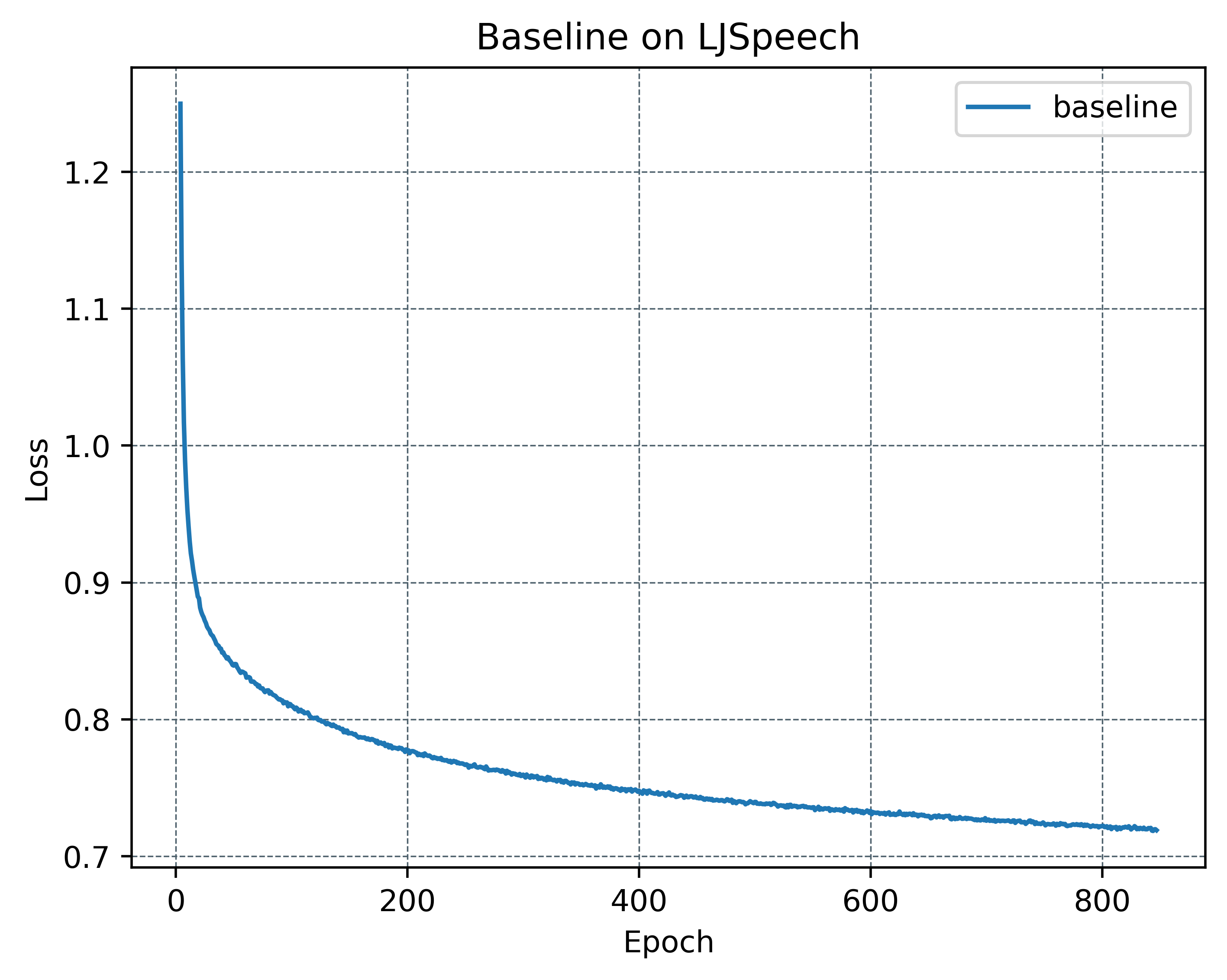}\label{sub1}
      }%scale是一个缩放
      \quad%空格
      \subfigure[Loss curve of our method]{
           \includegraphics[scale=0.38]{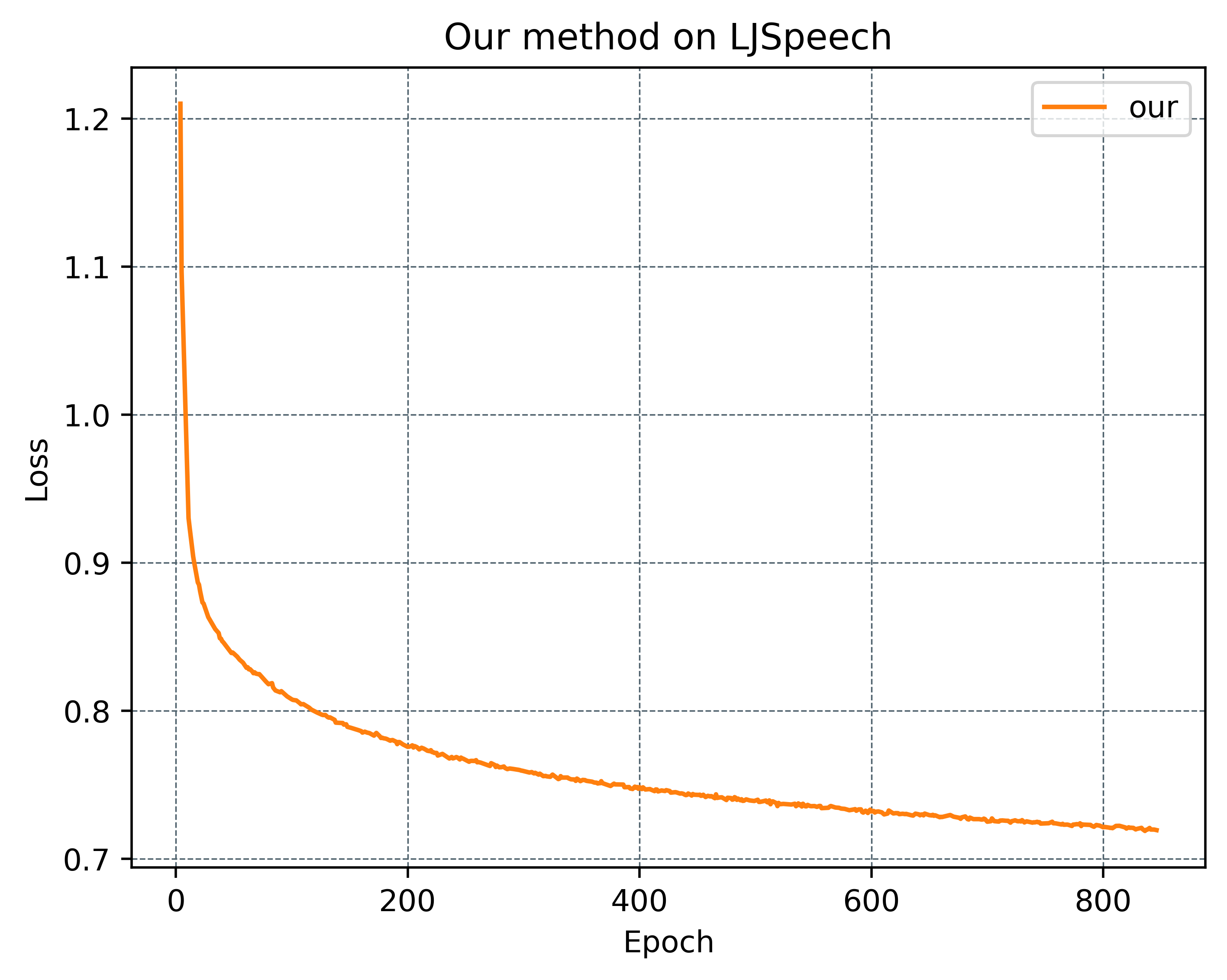}\label{sub2}
      }   
      \quad%空格
      \subfigure[Comparison of Training Speed]{
           \includegraphics[scale=0.38]{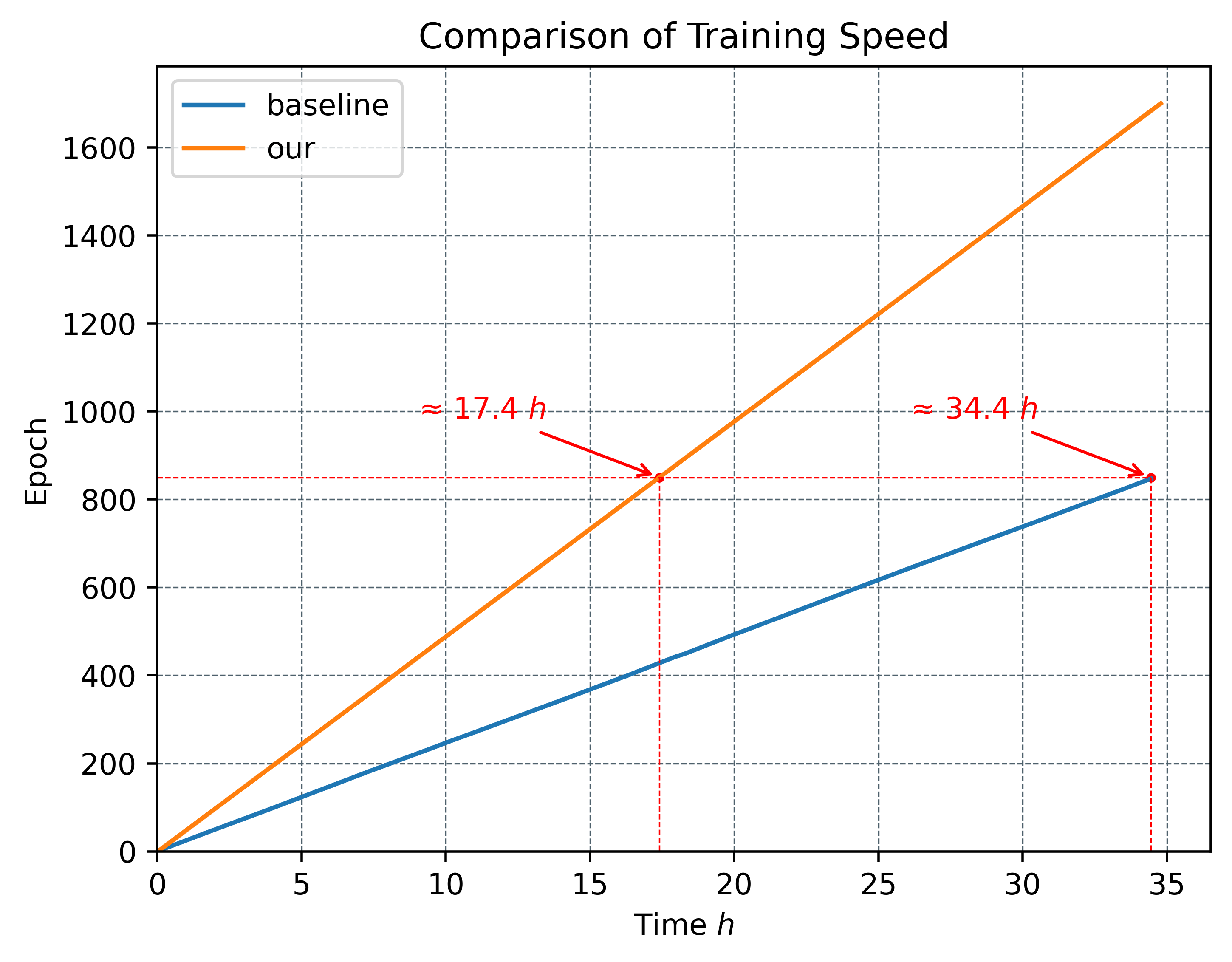}\label{sub3}
      } 
%我们的方法与基线模型的训练速度对比实验。(a)和(b)显示，在相同训练epoch内，我们的方法可以达到与基线模型几乎相同的训练效果。(c)显示在训练相同的epoch内，我们的训练速度快了将近一倍。
      \caption{We conducted comparative experiments on the training speed of our method versus the baseline model. Figures (a) and (b) show that within the same number of training epochs, our method can achieve nearly the same performance as the baseline model. Figure (c) shows that, within the same number of epochs, our training speed is nearly twice as fast.}
      \label{speed}
\end{figure*}

\subsection{Experimental Results}

%Directional Patch Interaction can train faster
\subsubsection{\textbf{Directional Patch Interaction enables fast training}}

%我们对比了DPI-TTS和基线Dex-TTS的训练速度。结果如图\ref{speed}所示。
%(a)(b)分别是DPI-TTS和基线在训练850个epoch的loss曲线，(c)是在相同训练时间内，DPI-TTS和基线的epoch数。
We compared the training speed of DPI-TTS with the baseline Dex-TTS, as shown in Figure \ref{speed}. Subfigures (a) and (b) display the loss curves of the baseline and DPI-TTS over 850 training epochs, respectively, while subfigure (c) illustrates the number of epochs completed by DPI-TTS and the baseline within the same training duration.

%实验结果表明，在相同epoch内，our method与基线相比没有精度损失，在相同时间内，DPI-TTS多训练了近一倍的epoch。
%这表明我们的方法在不损失训练效果的情况下，训练速度上快了接近一倍。
Experimental results indicate that, compared to the baseline, our method maintains accuracy within the same number of epochs. In the same time frame, DPI-TTS trains nearly twice as many epochs. 
This indicates that our method boosts the training speed nearly 2 times without compromising performance.

\subsubsection{\textbf{Fast training does not degrade performance}}
%我们通过对比WER和MOS-N来证明Directional Patch Interaction在保证语音时序一致性的有效性，对比COS来验证Style Temporal Modeling的有效性。在在单说话人数据集和多说话人数据集上的实验结果如表\ref{res}.
We demonstrate the effectiveness of Directional Patch Interaction in maintaining speech temporal consistency by comparing WER and MOS-N. The effectiveness of Style Temporal Modeling is validated through comparisons of COS. Experimental results on both single-speaker and multi-speaker datasets are presented in Table \ref{res}.

%DPI-TTS在WER和MOS-N的结果均优于基线。
%我们认为，让每个mel patch按照声学属性与前一阵和低频部分交互，而非整个频谱，有利于增强局部的建模细节，有效表现出差异比如“能量在高低频的不同”，从而提高了自然度。。
%同时，让每个patch依据声学属性顺序生成，在计算过程中强制增强了时序信息建模，从而加强了每一帧之间的连贯性。
DPI-TTS outperforms the baseline in both WER and MOS-N. 
This improvement is achieved by allowing each Mel spectrogram patch to interact with preceding frames and low-frequency components based on acoustic properties rather than considering the entire spectrum. 
This focused interaction enhances the modeling of local details and captures subtle differences, such as variations in energy between high and low frequencies, thereby improving the naturalness of the generated speech.
Moreover, the sequential generation of each patch, guided by acoustic properties, enforces effective temporal information modeling. 
This process strengthens the coherence between frames, resulting in more fluent and consistent TTS.

%DPI-TTS与在COS上表现优于基线。
%我们分析，风格时序性建模以一种细粒度的方式让patch完整的看到风格信息，同时以时间为顺序对同一时刻的所有patch为单位与风格信息融合有利于提升高低频风格表现的一致性，最终优化了Directional Patch Interaction下的风格表现。

DPI-TTS outperforms the baseline in COS. Our analysis suggests that fine-grained style temporal modeling enables each patch to fully incorporate style information. By sequentially integrating all patches with style information at each time point, DPI-TTS enhances the consistency of style representation across both high and low frequencies.

\begin{table}[t]
\centering
\caption{Comparison Of Results On LJSpeech And VCTK}
\label{res}
\begin{tabular}{lcccc}
\hline
Model         & WER ↓               & COS ↑                 & MOS-N ↑              & MOS-S ↑             \\ \hline
GT            & 6.56/6.23          & -                    & 4.64/4.52          & -                  \\
VITS          & 8.12/18.23         & 91.22/77.48          & 4.18/4.23          & 4.09/3.33          \\
Grad-TTS      & 7.70/17.84         & 91.37/79.92          & 4.27/4.28          & 4.14/3.47          \\
Dex-TTS       & 6.55/7.85          & 91.75/85.31          & 4.37/4.33          & 4.21/3.88          \\ \hline
DPI-TTS(ours) & \textbf{6.57/7.83} & \textbf{91.83/85.38} & \textbf{4.41/4.38} & \textbf{4.28/3.92} \\ \hline
\multicolumn{4}{l}{Left: LJSpeech results / Right: VCTK results.}
\end{tabular}
\end{table}

\begin{table}[t]
\centering
\caption{Directional Patch Interaction Ablation Study Results}
\label{ablation1}
\begin{tabular}{lcccc}
\hline
\multicolumn{1}{c}{Model} & WER ↓          & COS ↑            & MOS-N ↑         & MOS-S ↑         \\ \hline
DPI-TTS                   & \textbf{6.57} & \textbf{91.83} & \textbf{4.41} & \textbf{4.28} \\
$ph,p,h$                  & 6.97          & 90.79          & 4.24          & 4.13          \\
$p,pl$                    & 7.02          & 90.27          & 4.21          & 4.11          \\
$ph,p$                    & 7.13          & 90.18          & 4.19          & 4.07          \\
$h,nh,n$                  & 7.89          & 89.97          & 4.07          & 3.97          \\
$l,nl,n$                  & 7.93          & 90.01          & 4.11          & 4.09          \\
$n,nl$                    & 8.84          & 89.89          & 4.01          & 3.92          \\
$n,nh$                    & 8.72          & 89.46          & 3.98          & 3.87          \\ \hline
\end{tabular}
\end{table}

\begin{table}[t]
\centering
\caption{Other Ablation Study Results}
\label{ablation2}
\begin{tabular}{lcccc}
\hline
Model        & WER ↓         & COS ↑          & MOS-N ↑       & MOS-S ↑       \\ \hline
DPI-TTS      & \textbf{7.83} & \textbf{85.38} & \textbf{4.38} & \textbf{3.92} \\
w/o stm      & 8.34          & 83.53          & 4.16          & 3.72          \\
w/o tp       & 8.29          & 84.39          & 4.02          & 3.74          \\
w/o fp       & 9.02          & 84.96          & 3.95          & 3.75          \\
w/o tp \& fp & 9.83          & 83.87          & 3.76          & 3.78          \\ \hline
\multicolumn{4}{l}{w/o: without, stm: style temporal modeling}\\
\multicolumn{4}{l}{tp: time pos, fp: frequency pos, \&: and}
\end{tabular}
\end{table}

\subsection{Ablation Studies}
\subsubsection{Directional Patch Interaction}
%我们针对根据声学属性设计Directional Patch Interaction在LJSpeech数据集上进行消融实验。
%我们对所有相对合理的组合进行了评测。
%结果如表\ref{ablation1}所示。
%其中，$n$表示后一帧，$p$表示前一帧，,$l$表示低频,$h$表示高频，$ph$表示前一帧的高频部分，以此类推。

%结果表明，在违背了从低频到高频、从前一阵到后一帧的顺序进行推理时，WER、COS和MOS-N的结果都有不同程度的下降。
%这表明了依据语音物理特性进行推理的科学性和有效性。
We conducted ablation experiments on the LJSpeech dataset to evaluate the Directional Patch Interaction, which is designed based on acoustic properties. We assessed all reasonable combinations, and the results are presented in Table \ref{ablation1}. In this table, the following notations are used: \( n \) denotes the next frame, \( p \) denotes the previous frame, \( l \) represents low frequency, \( h \) represents high frequency, and \( ph \) indicates the high-frequency part of the previous frame, among others.
 
The results indicate that deviations from the low-to-high frequency sequence and the previous-to-next frame order lead to declines in WER, COS, and MOS-N scores to varying degrees. This finding underscores the necessity and effectiveness of inference based on acoustic properties. Adhering to these constraints ensures more accurate and reliable modeling of speech features.
\subsubsection{Other}
%我们对位置信息和时序风格建模在VCTK数据集上进行消融实验，结果如表\ref{ablation2}所示。

%当采用常规风格建模方法时，COS和MOS-S都有明显下降，这证明了style temporal modeling的方法有效性。

%在都取消和只取消其中一个位置信息的情况下，所有的指标结果都有不同程度的下降。
%我们认为，有向推理下的patch由于进行的是局部计算，添加这种全局的位置信息作为引导，能一定程度上提升模型结果的稳定性，并且添加位置信息有利于提升模型处理不同文本的能力。
We performed ablation experiments on ``positional information'' and ``style temporal modeling'' using the VCTK dataset. The results are presented in Table \ref{ablation2}.

When conventional style modeling methods are used, COS and MOS-S show significant declines, highlighting the effectiveness of style temporal modeling. Removing either type of positional information, or both, leads to varying degrees of decline across all metrics. We believe that, since Directional Patch Interaction involves local calculations, incorporating global positional information as guidance can enhance the stability of the model's results. Additionally, including positional information improves the model's ability to handle diverse texts.

\section{Conclusion}
This paper proposes a novel method called Directional Patch Interaction for Text-to-Speech (DPI-TTS). By integrating specific acoustic processing into the diffusion framework, DPI-TTS addresses the limitations of existing DiT speech models that treat the Mel spectrogram as a general image. The model employs frame-by-frame inference from low to high frequencies, enhancing the naturalness of synthetic speech while maintaining high training efficiency without compromising accuracy. 
Additionally, fine-grained style temporal modeling further improves speaker style similarity and sets DPI-TTS apart from traditional methods. Experimental results validate the effectiveness of the proposed method, demonstrating significant improvements in naturalness and style consistency. DPI-TTS offers a promising approach for transformer-based speech synthesis, contributing valuable insights and potential advancements to the field.

%\section*{Acknowledgment}

\balance
\bibliographystyle{IEEEtran}\
\bibliography{reference}

% Generated by IEEEtran.bst, version: 1.14 (2015/08/26)
\begin{thebibliography}{10}
\providecommand{\url}[1]{#1}
\csname url@samestyle\endcsname
\providecommand{\newblock}{\relax}
\providecommand{\bibinfo}[2]{#2}
\providecommand{\BIBentrySTDinterwordspacing}{\spaceskip=0pt\relax}
\providecommand{\BIBentryALTinterwordstretchfactor}{4}
\providecommand{\BIBentryALTinterwordspacing}{\spaceskip=\fontdimen2\font plus
\BIBentryALTinterwordstretchfactor\fontdimen3\font minus \fontdimen4\font\relax}
\providecommand{\BIBforeignlanguage}[2]{{%
\expandafter\ifx\csname l@#1\endcsname\relax
\typeout{** WARNING: IEEEtran.bst: No hyphenation pattern has been}%
\typeout{** loaded for the language `#1'. Using the pattern for}%
\typeout{** the default language instead.}%
\else
\language=\csname l@#1\endcsname
\fi
#2}}
\providecommand{\BIBdecl}{\relax}
\BIBdecl

\bibitem{shen2018natural}
J.~Shen, R.~Pang, R.~J. Weiss, M.~Schuster, N.~Jaitly, Z.~Yang, Z.~Chen, Y.~Zhang, Y.~Wang, R.~Skerrv-Ryan \emph{et~al.}, ``Natural tts synthesis by conditioning wavenet on mel spectrogram predictions,'' in \emph{2018 IEEE international conference on acoustics, speech and signal processing (ICASSP)}.\hskip 1em plus 0.5em minus 0.4em\relax IEEE, 2018, pp. 4779--4783.

\bibitem{ren2019fastspeech}
Y.~Ren, Y.~Ruan, X.~Tan, T.~Qin, S.~Zhao, Z.~Zhao, and T.-Y. Liu, ``Fastspeech: Fast, robust and controllable text to speech,'' \emph{Advances in neural information processing systems}, vol.~32, 2019.

\bibitem{ren2020fastspeech}
Y.~Ren, C.~Hu, X.~Tan, T.~Qin, S.~Zhao, Z.~Zhao, and T.-Y. Liu, ``Fastspeech 2: Fast and high-quality end-to-end text to speech,'' \emph{arXiv preprint arXiv:2006.04558}, 2020.

\bibitem{ronneberger2015u}
O.~Ronneberger, P.~Fischer, and T.~Brox, ``U-net: Convolutional networks for biomedical image segmentation,'' in \emph{Medical image computing and computer-assisted intervention--MICCAI 2015: 18th international conference, Munich, Germany, October 5-9, 2015, proceedings, part III 18}.\hskip 1em plus 0.5em minus 0.4em\relax Springer, 2015, pp. 234--241.

\bibitem{jeong2021diff}
M.~Jeong, H.~Kim, S.~J. Cheon, B.~J. Choi, and N.~S. Kim, ``Diff-tts: A denoising diffusion model for text-to-speech,'' \emph{arXiv preprint arXiv:2104.01409}, 2021.

\bibitem{popov2021grad}
V.~Popov, I.~Vovk, V.~Gogoryan, T.~Sadekova, and M.~Kudinov, ``Grad-tts: A diffusion probabilistic model for text-to-speech,'' in \emph{International Conference on Machine Learning}.\hskip 1em plus 0.5em minus 0.4em\relax PMLR, 2021, pp. 8599--8608.

\bibitem{huang2022prodiff}
R.~Huang, Z.~Zhao, H.~Liu, J.~Liu, C.~Cui, and Y.~Ren, ``Prodiff: Progressive fast diffusion model for high-quality text-to-speech,'' in \emph{Proceedings of the 30th ACM International Conference on Multimedia}, 2022, pp. 2595--2605.

\bibitem{ye2023comospeech}
Z.~Ye, W.~Xue, X.~Tan, J.~Chen, Q.~Liu, and Y.~Guo, ``Comospeech: One-step speech and singing voice synthesis via consistency model,'' in \emph{Proceedings of the 31st ACM International Conference on Multimedia}, 2023, pp. 1831--1839.

\bibitem{tan2024naturalspeech}
X.~Tan, J.~Chen, H.~Liu, J.~Cong, C.~Zhang, Y.~Liu, X.~Wang, Y.~Leng, Y.~Yi, L.~He \emph{et~al.}, ``Naturalspeech: End-to-end text-to-speech synthesis with human-level quality,'' \emph{IEEE Transactions on Pattern Analysis and Machine Intelligence}, 2024.

\bibitem{shen2023naturalspeech}
K.~Shen, Z.~Ju, X.~Tan, Y.~Liu, Y.~Leng, L.~He, T.~Qin, S.~Zhao, and J.~Bian, ``Naturalspeech 2: Latent diffusion models are natural and zero-shot speech and singing synthesizers,'' \emph{arXiv preprint arXiv:2304.09116}, 2023.

\bibitem{kong2020diffwave}
Z.~Kong, W.~Ping, J.~Huang, K.~Zhao, and B.~Catanzaro, ``Diffwave: A versatile diffusion model for audio synthesis,'' \emph{arXiv preprint arXiv:2009.09761}, 2020.

\bibitem{huang2022fastdiff}
R.~Huang, M.~W. Lam, J.~Wang, D.~Su, D.~Yu, Y.~Ren, and Z.~Zhao, ``Fastdiff: A fast conditional diffusion model for high-quality speech synthesis,'' \emph{arXiv preprint arXiv:2204.09934}, 2022.

\bibitem{chen2022infergrad}
Z.~Chen, X.~Tan, K.~Wang, S.~Pan, D.~Mandic, L.~He, and S.~Zhao, ``Infergrad: Improving diffusion models for vocoder by considering inference in training,'' in \emph{ICASSP 2022-2022 IEEE International Conference on Acoustics, Speech and Signal Processing (ICASSP)}.\hskip 1em plus 0.5em minus 0.4em\relax IEEE, 2022, pp. 8432--8436.

\bibitem{koizumi2022specgrad}
Y.~Koizumi, H.~Zen, K.~Yatabe, N.~Chen, and M.~Bacchiani, ``Specgrad: Diffusion probabilistic model based neural vocoder with adaptive noise spectral shaping,'' \emph{arXiv preprint arXiv:2203.16749}, 2022.

\bibitem{lee2021priorgrad}
S.-g. Lee, H.~Kim, C.~Shin, X.~Tan, C.~Liu, Q.~Meng, T.~Qin, W.~Chen, S.~Yoon, and T.-Y. Liu, ``Priorgrad: Improving conditional denoising diffusion models with data-dependent adaptive prior,'' \emph{arXiv preprint arXiv:2106.06406}, 2021.

\bibitem{peebles2023scalable}
W.~Peebles and S.~Xie, ``Scalable diffusion models with transformers,'' in \emph{Proceedings of the IEEE/CVF International Conference on Computer Vision}, 2023, pp. 4195--4205.

\bibitem{vaswani2017attention}
A.~Vaswani, ``Attention is all you need,'' \emph{Advances in Neural Information Processing Systems}, 2017.

\bibitem{jing2023u}
X.~Jing, Y.~Chang, Z.~Yang, J.~Xie, A.~Triantafyllopoulos, and B.~W. Schuller, ``U-dit tts: U-diffusion vision transformer for text-to-speech,'' in \emph{Speech Communication; 15th ITG Conference}.\hskip 1em plus 0.5em minus 0.4em\relax VDE, 2023, pp. 56--60.

\bibitem{ljspeech17}
K.~Ito, ``The lj speech dataset,'' \url{https://keithito.com/LJ-Speech-Dataset/}, 2017.

\bibitem{lee2024ditto}
K.~Lee, D.~W. Kim, J.~Kim, and J.~Cho, ``Ditto-tts: Efficient and scalable zero-shot text-to-speech with diffusion transformer,'' \emph{arXiv preprint arXiv:2406.11427}, 2024.

\bibitem{park2024dex}
H.~J. Park, J.~S. Kim, W.~Shin, and S.~W. Han, ``Dex-tts: Diffusion-based expressive text-to-speech with style modeling on time variability,'' \emph{arXiv preprint arXiv:2406.19135}, 2024.

\bibitem{hu2024mel}
G.~Hu, H.~Tan, and R.~Li, ``A mel spectrogram enhancement paradigm based on cwt in speech synthesis,'' \emph{arXiv preprint arXiv:2406.12164}, 2024.

\bibitem{zhou2024mel}
R.~Zhou, X.~Li, Y.~Fang, and X.~Li, ``Mel-fullsubnet: Mel-spectrogram enhancement for improving both speech quality and asr,'' \emph{arXiv preprint arXiv:2402.13511}, 2024.

\bibitem{xu2024convconcatnet}
X.~Xu, B.~Wang, Y.~Yan, H.~Zhu, Z.~Zhang, X.~Wu, and J.~Chen, ``Convconcatnet: a deep convolutional neural network to reconstruct mel spectrogram from the eeg,'' \emph{arXiv preprint arXiv:2401.04965}, 2024.

\bibitem{tian2023diffusion}
Y.~Tian, W.~Liu, and T.~Lee, ``Diffusion-based mel-spectrogram enhancement for personalized speech synthesis with found data,'' in \emph{2023 IEEE Automatic Speech Recognition and Understanding Workshop (ASRU)}.\hskip 1em plus 0.5em minus 0.4em\relax IEEE, 2023, pp. 1--7.

\bibitem{qian2022cmelgan}
T.~Qian, J.~Kaunismaa, and T.~Chung, ``Cmelgan: An efficient conditional generative model based on mel spectrograms,'' \emph{arXiv preprint arXiv:2205.07319}, 2022.

\bibitem{su2024roformer}
J.~Su, M.~Ahmed, Y.~Lu, S.~Pan, W.~Bo, and Y.~Liu, ``Roformer: Enhanced transformer with rotary position embedding,'' \emph{Neurocomputing}, vol. 568, p. 127063, 2024.

\bibitem{kim2021conditional}
J.~Kim, J.~Kong, and J.~Son, ``Conditional variational autoencoder with adversarial learning for end-to-end text-to-speech,'' in \emph{International Conference on Machine Learning}.\hskip 1em plus 0.5em minus 0.4em\relax PMLR, 2021, pp. 5530--5540.

\bibitem{Veaux2017CSTRVC}
C.~Veaux, J.~Yamagishi, and K.~MacDonald, ``Cstr vctk corpus: English multi-speaker corpus for cstr voice cloning toolkit,'' 2017.

\bibitem{kingma2013auto}
D.~P. Kingma, ``Auto-encoding variational bayes,'' \emph{arXiv preprint arXiv:1312.6114}, 2013.

\bibitem{rezende2015variational}
D.~Rezende and S.~Mohamed, ``Variational inference with normalizing flows,'' in \emph{International conference on machine learning}.\hskip 1em plus 0.5em minus 0.4em\relax PMLR, 2015, pp. 1530--1538.

\end{thebibliography}

\end{document}